\documentclass[tighten,twocolumn]{aastex631}
\usepackage[]{graphicx}
\usepackage{amsmath}
\usepackage[normalem]{ulem}
\newcommand{\ndet}{\mathrm{N_{det}}}
\newcommand{\ntmpl}{\mathrm{N_{tmpl}}}
\newcommand{\nb}{\mathrm{N_{b}}}
\newcommand{\nf}{\mathrm{N_{F}}}
\newcommand{\npix}{\mathrm{N_{pix}}}
\newcommand{\nchan}{\mathrm{N_{chan}}}
\newcommand{\tmin}{\mathrm{\tau_{min}}}
\newcommand{\tb}{\mathrm{\tau_{b}}}

\shortauthors{Kerr et al.}
\shorttitle{Real-time $\gamma$-ray Transient Detection}

\begin{document}

\title{Real-time Likelihood Methods for Improved $\gamma$-ray
Transient Detection and Localization}

\author[0000-0002-0893-4073]{M.~Kerr}
\author[0000-0001-9322-6153]{W.~Duvall}
\author{W.~N.~Johnson}
\author[0000-0003-4859-1711]{R.~S.~Woolf}
\author[0000-0002-0586-193X]{J.~E.~Grove}
\author{H.~Kim}
\affiliation{Space Science Division, Naval Research Laboratory, Washington, DC 20375--5352, USA}
\correspondingauthor{M.~Kerr}
\email{matthew.kerr@nrl.navy.mil}

\begin{abstract}We present a maximum likelihood (ML) algorithm that
  is fast enough to detect $\gamma$-ray transients in real time on
  low-performance processors often used for space applications.  We
  validate the routine with simulations and find that, relative to
  algorithms based on excess counts, the ML method is nearly twice as
  sensitive, allowing detection of 240--280\% more short
  $\gamma$-ray bursts. We characterize a reference implementation of
  the code, estimating its computational complexity and benchmarking
  it on a range of processors.  We exercise the reference
  implementation on archival data from the \textit{Fermi} Gamma-ray
  Burst Monitor (GBM), verifying the sensitivity improvements.  In
  particular, we show that the ML algorithm would have detected
  GRB~170817A even if it had been nearly four times fainter.  We
  present an ad hoc but effective scheme for discriminating transients
  associated with background variations.  We show that the on-board
  localizations generated by ML are accurate, but that refined
  off-line localizations require a detector response matrix with about
  ten times finer resolution than is current practice.  Increasing the
  resolution of the GBM response matrix could substantially reduce the
  few-degree systematic uncertainty observed in the localizations of
  bright bursts.
\vspace{1cm} \end{abstract}


\section{Introduction}
\label{sec:intro}

The prompt emission from $\gamma$-ray bursts (GRBs) is typically
concentrated between 100\,keV and 1\,MeV
\citep{vonKienlin20,Poolakkil21} and varies in duration from $<$2\,s
\citep[short GRBs,][]{Kouveliotou93} to minutes (long GRBs), to
hours \citep[rare ultra-long GRBs,][]{Levan14}.  These soft
$\gamma$ rays primarily Compton scatter,
limiting detection technologies. Wideband Compton telescopes
reconstruct the photon interaction and provide a wide field-of-view
and modest effective area, but they require very high-resolution,
expensive detector elements and readout systems
\citep[e.g.][]{Tomsick19,McEnery19}.  Coded masks enable indirect
imaging up to 100--200\,keV (when the mask becomes transparent) with
modest field-of-view and angular resolution: the Burst Alert
Telescope is capable of providing GRB localizations to $<$1'
\citep{Barthelmy05} in a 30$^\circ$ field-of-view.  Laue lenses allow
direct focusing via coherent Bragg scattering \citep{Frontera10}, but
their bulk, cost, and narrow field-of-view are impractical for all-sky
monitoring.  

Scintillating crystals provide effective stopping power over a wide
energy range and can be read out with simple electronics.
Doped NaI and CsI can be grown to large sizes, offering exceptional
sensitivity-to-cost ratio, and detectors based on these crystals have
been widely used for $\gamma$-ray transient detection.  The Vela
nuclear test monitoring system used CsI detectors in its serendipitous
discovery of GRBs \citep{Klebesadel73}.   The pioneering
\citep{Meegan92} Burst and Transient Source Experiment (BATSE) on the
\textit{Compton} Gamma-Ray Observatory used very large NaI
scintillators that were viewed from the side with photomultiplier
tubes.  The Gamma-ray Burst Monitor (GBM) on the \textit{Fermi}
Gamma-Ray Space Telescope uses much smaller NaI pucks with Be windows read
from the large face to maximize light yield, enabling a particularly
low ($\sim$10\,keV) threshold \citep{Meegan09}.  Future experiments
like Glowbug \citep{Woolf22} and Starburst \citep{Kocevski2022StarBurst}
will use large, thin crystals read out from
the edge with silicon photomultipliers \citep[SiPMs,][]{Mitchell21} to
achieve very high sensitivity in a compact, low-voltage, low-cost
design.

Scintillators have fast scintillation and readout times, typically
$\lesssim$1\,$\mu$s, enabling photon counting but
providing no position information.  Instead, comparing rates
between two detectors with differing incidence angles constrains the
source position to a fuzzy great circle.  Additional detector pairs
provide further constraints and reduce the uncertainty, and many pairs
yield a localization that is well-described as a single point with a
gaussian-distributed uncertainty.  Increasing the signal-to-background
ratio further reduces the uncertainty, allowing for designs with a few
large facets (Glowbug) or many smaller facets (GBM) to produce
comparable localizing power.

Scintillators also provide no intrinsic discrimination\footnote{Phoswich designs can provide background
discrimination but increase cost and complexity.} of background
events, e.g. from energetic particles trapped in
earth's magnetosphere or $\gamma$ rays produced by internal
radioactivation.
Time-dependent background variations can mimic a
$\gamma$-ray transient, so a detector system must first
recognize it, i.e. trigger.  Triggers often
initiate resource-intensive processes, such as re-pointing the
spacecraft or downlinking a large volume of data for offline analysis.
Thus, trigger criteria and algorithms must be carefully tuned to
provide sensitivity to transients while minimizing false positives.
GBM provides one example of an effective on-board trigger scheme
\citep{Meegan09}: it autonomously measures the background rate in each
NaI detector, averages excess counts over a range of timescales, and
triggers when two or more detectors register a $>$4.5$\sigma$ excess.

Large SiPM-read crystals, the availability of cheap commercial
SmallSat spacecraft buses, and rideshare launch opportunities mean it
is now possible to launch $\gamma$-ray transient detectors with the
performance of heritage instruments at a fraction of the cost.  Such
detectors will have limited telemetry bandwidth, so a sensitive
trigger is required to maximize the scientific merit of the data that
are downlinked.  For example, the short GRBs accompanying neutron
star mergers are expected to be faint because future
gravitational wave detectors will detect more distant mergers and
because the jetted emission will not generally point towards earth.
GRB~170817A was exceptionally close but was only 7\% over the GBM
trigger threshold \citep{Goldstein17}.

In this work, we propose to improve trigger sensitivity and efficiency
by using maximum likelihood (ML) for the on-board detection and
localization of GRBs and other $\gamma$-ray transients.  ML methods
are already used for offline analysis of GRB spectra and positions
\citep{Berlato19,Goldstein20} and for post facto detection
\citep{Blackburn15,Kocevski18}.  We show that such methods are capable of running
in real-time on the low-power, low-performance processors typically
used for space applications and that they can detect transients that
are half as bright as those detectable with rate-based triggers.
Moreover, ML links detection and localization, so a coarse
position is available immediately upon detection, enabling very
low-latency alerts.  The same framework can be used to rapidly refine
the localization.  Finally, ML methods can be tuned to different
source classes, including backgrounds, allowing efficient filtering of
desired transients.

In the next section, we develop the formalism for ML detection and
localization with scintillator detectors and show that some simple
approximations yield fast but reliable estimates of transient
significance.  In \$\ref{sec:sims}, we use Monte Carlo simulations of
a GBM-like dataset to validate the ML estimators for detection and
localization and to estimate the threshold transient flux, allowing
an estimate of the relative sensitivity.  We describe the
computational performance of a reference implementation in
\S\ref{sec:impl} in theory and as benchmarked on relevant hardware.
We characterize the algorithmic performance in \S\ref{sec:realworld},
demonstrating both sensitivity to \textit{bona fide} GRBs and an
effective method of reducing false triggers from particle backgrounds.
We consider the accuracy of both coarse and follow-up localization in
\S\ref{sec:localization}, finding that the reliability depends
strongly on the spatial resolution of the detector response matrix.
Finally, we conclude with a summary of the results and a discussion of
potential new applications enabled by the method in
\S\ref{sec:discussion}.

\section{Fast likelihood methods}
\label{sec:like}

$\gamma$-ray/$e^{-}$ interactions in a scintillator release a pulse of
optical light that is collected, amplified, and measured.  For a
thick scintillator, the incident energy $E'$ is entirely
converted to a pulse height $E$ proportional to $E'$
with a spectral resolution of order 5\%.  For thin
crystals/higher energies, scattered photons and electrons may
escape, yielding $E$ possibly much less than $E'$.  The conversion
efficiency (effective area) also drops with
increasing energy until pair production begins.

The effective area and the energy redistribution can be
encapsulated in a {response matrix} (RM) that converts $\gamma$ rays
from a distant source incident at angle $\Omega$ with spectral flux density $F(E')$
($\mathrm{ph}\,\mathrm{keV}^{-1}\,\mathrm{s}^{-1}\,\mathrm{cm}^{-2}$)
into an observed counts spectrum 
$N(E)$
($\mathrm{ph}\,\mathrm{keV}^{-1}\,\mathrm{s}^{-1}$):
\begin{equation}
  N(E) = \int_0^\infty dE' R(E,E',\Omega) F(E',\Omega).
\end{equation}

The RM is typically estimated with Monte Carlo
simulations and particle transport codes and tabulated as a matrix,
$R_{lijq}$: the response to
a pixel centered on $\Omega_l$, for a detector with index $i$, averaged
over observed/measured
energy channel centered on $E_j$, averaged over incident energy
channel centered on $E_q$.  $R$ typically depends on time, and a
spacecraft pointing history relates source coordinates to instrument
coordinates.  We adopt a source model with a background rate
($b_{ij}$) for each
detector element ($i$) and energy channel ($j$) and a point
source at position $\Omega_l$ with spectrum $F_k(E)$ ($k$ just labels
this spectral shape).  
The predicted
counts per unit time are then
\begin{align}
  \lambda_{klij} = &\ b_{ij} + \alpha_{kl} \sum_q R_{lijq} F_k(E_q) \\
               \equiv&\ b_{ij} + \alpha_{kl} F_{klij}.
\end{align}
The second line re-expresses the convolution of the source spectrum
with $R$ as a fiducial ``template'' $F$ for the observed
counts.  The fiducial template is scaled by the source amplitude
$\alpha_{kl}$ to produce the final counts prediction.  The observed
counts $n_{ij}$ follow a Poisson distribution, so an estimator for the
source amplitude $\hat{\alpha}_{kl}$ can be obtained by maximizing the
log likelihood
\begin{widetext}
\begin{equation}
  \log \mathcal{L}(\alpha_{kl}) = \sum_{i=1}^{\ndet} \sum_{j=1}^{\nchan}  n_{ij} \log \left( b_{ij} +
  \alpha_{kl} F_{klij}\right) - b_{ij} -\alpha_{kl} F_{klij}.
\end{equation}
Source significance can be estimated with the log likelihood ratio
test statistic $\mathrm{TS}\equiv2\times\left[\log
\mathcal{L}(\hat{\alpha_{kl}})-\log \mathcal{L}(0)\right]$.
\citet{Wilks38} shows that $\mathrm{TS}$ asymptotically follows a
$\chi^2_1$ distribution.  By appeal to the Cram\'{e}r-Rao bound, this
is also the most sensitive detection statistic.

For detection, we must scan over possible source positions and
spectral shapes, obtaining sets of $\alpha_{kl}$ and
$\mathrm{TS}(\alpha_{kl})$ for each time interval of interest.  These
evaluations are expensive, and a fast
method is critical for real-time applications.  We can focus on
transients that are near-threshold, because bright transients will
produce obvious signals.  Thus, we write
\begin{align}
  \log \mathcal{L}(\alpha_{kl}) =&\ \sum_{i=1}^{\ndet} \sum_{j=1}^{\nchan}  n_{ij} \log \left( 1 +
  \alpha_{kl} F_{klij}/b_{ij}\right) - b_{ij} -\alpha_{kl} F_{klij} +
  \mathrm{const.}\\
  \label{eq:logl}\equiv &\ \sum_{i=1}^{\ndet} \sum_{j=1}^{\nchan} \left[ n_{ij} \log \left( 1 +
  \alpha_{kl} t_{klij}\right)\right] - B -\alpha_{kl} F_{kl},
\end{align}
where we have defined $t$ as the ratio of source to background counts
  and have defined data-integrated quantities $B$ and $F_{kl}$.
Taylor expanding the logarithm yields
\begin{align}
  \log \mathcal{L}(\alpha_{kl}) = &\ \sum_{i=1}^{\ndet} \sum_{j=1}^{\nchan}\left[ n_{ij} \left(
  \alpha_{kl} t_{klij} - \frac{1}{2} (\alpha_{kl} t_{klij})^2 + \frac{1}{3}
  (\alpha_{kl} t_{klij})^3 + \dots \right)\right] - B -\alpha_{kl}
  F_{kl}\\
  \equiv &\ \alpha_{kl} \langle NT_{kl}\rangle - \frac{\alpha_{kl}^2}{2}\langle
  NT_{kl}^2\rangle +\frac{\alpha_{kl}^3}{3}\langle NT_{kl}^3\rangle +\dots -B
  -\alpha_{kl} F_{kl}\label{eq:nt_like},
\end{align}
\end{widetext}
where quantities like $\langle NT\rangle$ are moments of the data
(counts) weighted by $t$.  Differentiating and discarding remaining
terms $\mathcal{O}(\alpha^2)$ and higher yields the first-order
estimator
\begin{equation}
  \hat{\alpha}_{1kl} = \frac{\langle NT_{kl} \rangle-F_{kl}}{\langle
  NT_{kl}^2
  \rangle},
\end{equation}
with $F_l$ now denoting the sum of all predicted counts.  Inserting this estimator into the likelihood ratio test statistic
  and discarding terms $\mathcal{O}(\alpha^3)$ yields the simple
  first-order estimator
\begin{equation}
  \label{eq:ts1}
  \mathrm{TS}_1(\hat{\alpha}_{1kl}) = \hat{\alpha}_{1kl}^2 \langle
  NT_{kl}^2
  \rangle = \frac{\left( \langle NT_{kl}
  \rangle-F_{kl}\right)^2}{\langle NT_{kl}^2 \rangle}
\end{equation}
As we show below, it is helpful to consider a higher-order estimator
formed from $\hat{\alpha}_{1kl}$ and the $\mathcal{O}(\alpha^3)$
terms in Eq. \ref{eq:nt_like}:
\begin{equation}
  \label{eq:ts2}
  \mathrm{TS}_2(\hat{\alpha}_{1kl}) = \mathrm{TS}_1(\hat{\alpha}_{1kl})
  +\frac{2}{3}\hat{\alpha}_{1kl}^3\langle NT_{kl}^3\rangle
\end{equation}

This expression neglects higher-order corrections to
$\hat{\alpha}_{1kl}$, because (1) doing so requires choosing the
  correct solution to a quadratic equation and (2)
the larger error is in the evaluation of
$\mathrm{TS}$, so this simple correction is effective.
  
These expressions are summed over observed energy, but it is also
useful to consider the estimators for each energy channel.  Finally, we note that these expressions can be
  used to iteratively obtain the full maximum likelihood estimator
  $\hat{\alpha}_{kl}$ and hence the exact TS, which we denote
  $\mathrm{TS_e}(\hat{\alpha}_{kl})$.

Thus, a real-time, maximum likelihood based technique for transient
  detection (and localization) is as follows: (1) group events
  into time bins, e.g. 64\,ms, 2,048\,ms, \dots; (2) evaluate the
  approximate TS (TS$_1$ or TS$_2$) over a set of pixels corresponding to possible transient
  positions (unocculted space for GRBs, towards the nadir for
  terrestrial $\gamma$-ray flashes, etc.); (3) assess the significance
  of any large values to evaluate trigger criteria.  We discuss the
  practical and efficient implementation of this approach further in
  \S\ref{sec:impl}.  However, we first illustrate general
  properties of the ML estimator on somewhat idealized
  data and compare it to existing real-time detection schemes.

\begin{table}
\centering
\begin{tabular}{c | c | r | r | r | r }
  Channel   & Trigger & E$_\mathrm{min}$ & E$_\mathrm{max}$ & Bkg.
  Rate & Bkg. Std\\
            & Channel & (keV)           & (kev)            & (evt\,s$^{-1}$) & (evt\,$s^{-1}$) \\
\hline
\hline
  0   & 0 &30    & 50   & 161 & 20\\
  1   & 1 &50    & 82   & 117 & 7 \\
  2   & 1 &82    & 135  & 99  & 5 \\
  3   & 1 &135   & 223  & 73  & 6 \\
  4   & 1 &223   & 367  & 42  & 3 \\
  5   & 2 &367   & 606  & 26  & 2 \\
  6   & 2 &606   & 1000 & 51  & 18\\
  7   & -- &1000 & 2000 & 38  & 16 \\
\end{tabular} \caption{\label{tab:channels}Channel edges used in
  simulation and later GBM data analysis.  Note that the last channel
  includes overflow events.  The second column indicates which
  channels are used in the GBM-like counts excess trigger.  The
  background event rate, averaged over the 12 NaI detectors, is given
here for reference, as is the variation (standard deviation) over the detectors.}\end{table}

\section{Calibration and Sensitivity}
\label{sec:sims}
In order to determine the sensitivity of the likelihood method, it is
first necessary to determine the distribution of the likelihood test
statistic in the absence of a signal, and thus a false alarm rate for
any non-stationary signal (GRBs or other transients).  For a given
false alarm rate, the threshold can be mapped to a sensitivity in
terms of transient flux.  We carry out this procedure with
simulations.

\subsection{Simulation Setup}

Archival GBM data, with its high time resolution, is ideal for
validating burst-detection algorithms.  To evaluate performance with
known signal and background characterists, we here simulate
``GBM-like'' data.  First, from archival data we estimate the typical
background $b_{ai}$ for each NaI detector in 8 energy channels (Table
\ref{tab:channels}).  The variation in rates over detectors is modest except at the highest energies.

\begin{figure}
\centering
  \includegraphics[angle=0,width=0.98\linewidth]{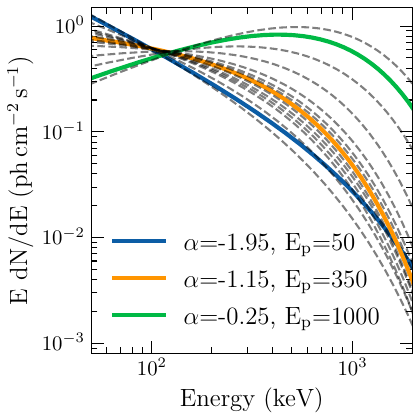}
\caption{\label{fig:models}Spectral models for simulation and
  detection of GRBs.  The colored lines represent three representative
  GRB spectra from \citet{Goldstein20} and form a basis for detection,
  while the 12 black dashed lines form a basis for simulation.}
\end{figure}

\citet[][hereafter G20]{Goldstein20} reported a set of three
representative ``Comptonized power law'' models,
\begin{equation}
  F(E) = N_0
\left(\frac{E}{100\,\mathrm{keV}}\right)^\alpha \exp\left[
  -(2+\alpha)\frac{E}{E_{\mathrm{p}}}\right].
\end{equation}
with parameters chosen to represent an unusually soft-spectrum GRB
($\alpha=-1.95$, $E_{\mathrm{p}}=50$\,keV), a typical GRB
($\alpha=-1.15$, $E_{\mathrm{p}}=350$\,keV), and an unusually
hard-spectrum GRB ($\alpha=-0.25$, $E_{\mathrm{p}}=1000$\,keV).  We
adopt these as our detection templates.  However, to simulate a
broader population, we have devised 12 ``basis'' spectra that do not
overlap the G20 templates and cover a range around them  (see Figure
\ref{fig:models}.)  They are concentrated most densely around the
``normal'' model, and by choosing randomly between them, we can
approximate drawing randomly from the population of GRBs.  (The
softest and hardest simulated spectra are even more extreme than G20
models, so in the simulations below we reduce the weight of these two
templates by 50\%.) The normalizations $N_0$ are chosen to give a
50--300\,keV flux of  1\,ph\,cm$^{-2}$\,s$^{-1}$.
We fold these basis spectral
models through the GBM RM, obtained from the files distributed with
\texttt{gbmrsp}\footnote{https://fermi.gsfc.nasa.gov/ssc/data/analysis/rmfit/}.
For consistency with the rest of our software, we resample the
provided 272 spatial pixels to a 482-pixel icosahedral 
tessellation of the sky.  The results predict the counts in 8 output
channels (see Table \ref{tab:channels} for boundaries) for each possible incidence direction.

To simulate a GRB---the alternative hypothesis---we choose one of the
12 spectral basis models and 482 incident directions at random, add
the source rate to the background, scale the predicted rates by the
duration $\delta t$, and draw Poisson random variables with the
resulting mean, producing an array of $\ndet\times\nchan$
($12\times8$) photon counts.  (We use the 12 NaI detectors but neglect
the 2 BGO detectors.)  The null hypothesis is obtained in the same way
but with zero source contribution.

\subsection{Calibrating the Test Statistic}

For each simulated data set, we evaluate the TS for the three G20
templates and for the 482 incident directions.  (We reiterate that the
simulation and detection templates are different.)  We first examined
the null hypothesis and found that the distribution of TS for a single
template/pixel follows the expected asymptotic distribution
($\chi^2_1$) almost perfectly, even for short time windows ($\delta
t=$64\,ms) with typically only a few counts in each channel.  However,
the detection statistic is the \textbf{maximum} TS over pixels and
templates, for which the distribution is not known.
Furthermore, the sample values of TS are
not independent because of finite angular resolution
and because the GRB templates are not orthogonal.  (The
extent to which the pixels over- or undersample the angular resolution
depends on the brightness of a given burst.  Thus the distribution of
the TS in the alternative hypothesis will in general be even more
complicated.)

All of this means that the significance of any apparently-large TS
must be calibrated with simulations.  Fortunately, because the null
distribution of TS does not depend on the time window $\delta t$, we
can calibrate TS universally simply by simulating many realization of
the null distribution ($\alpha\rightarrow0$) and calculating the
maximum TS over the pixels and templates, in this specific case,
$3\times482=1446$ values.

\begin{figure}
\centering
  \includegraphics[angle=0,width=0.98\linewidth]{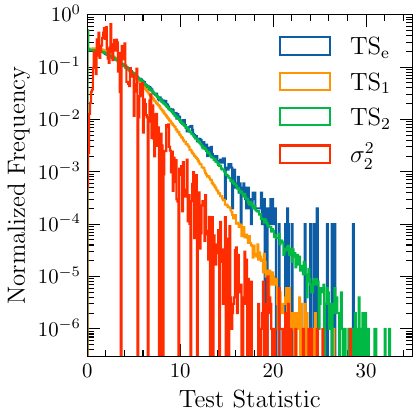}
\caption{\label{fig:tsnull}The maximum TS from a scan over 482
  possible incident directions and 3 trial template spectra from
  simulations of the null hypothesis, i.e. background only.
  $\sigma_2$ is the Poisson significance of the third highest rate
  excess in  the 12 detectors.  There are \textbf{$10^7$} realizations
  of TS$_1$, TS$_2$, and $\sigma_2$ and \textbf{$10^4$} of the more
  computationally expensive $\mathrm{TS_e}$.}
\end{figure}

The results for many simulations are shown in Figure \ref{fig:tsnull}.
Because the exact TS is expensive to evaluate, we have carried out
extensive simulations (N=$10^7$) only for TS$_1$ and TS$_2$.  We
further calculate $\sigma_2$, a detection statistic similar to the one
operating on GBM.  Of the 12 detectors, it is the second-highest
excess rate expressed in ``sigma'' units: $(c-b)/\sqrt{b}$, with $c$
the observed counts and $b$ the expected background.   The counts are
chosen in three coarse channels (see Table \ref{tab:channels}) that
approximately match the bounds used for the on-board GBM trigger, and
the the maximum excess from these three coarse channels is $\sigma_2$.

\begin{figure}
\centering
  \includegraphics[angle=0,width=0.98\linewidth]{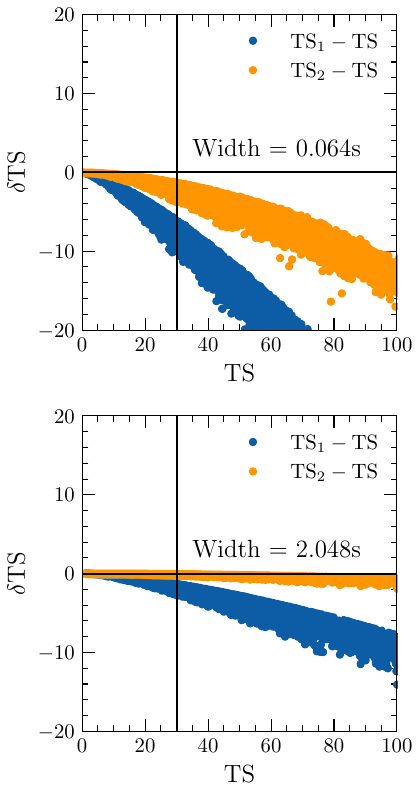}
\caption{\label{fig:tscomp}A comparison of exact vs. approximate
  evaluation of the test statistic for two characteristic time
  windows.  Both estimators agree well for the longer
  window---which has more photons---while only TS$_2$ is a sufficient approximation for short search windows.}
\end{figure}

It is clear (Figure \ref{fig:tsnull}) that the approximate estimators TS$_1$ and TS$_2$ undershoot the exact $\mathrm{TS_e}$.  For the purposes of detection,
the TS estimator needs only be accurate up to a typical threshold of $\sim$30.  To
gauge the differences more accurately, we simulate bursts with a
small, real signal, in order to shift the distribution to higher
values of TS.  The results for $\delta t=64$\,ms and
2048\,ms are shown in Figure \ref{fig:tscomp}.  Both approximate and
exact methods work for the longer window, but in
the shorter window, where there are fewer photons, the linear
approximation TS$_1$ is lower than the exact value by up to 30\%.  In
this case, one must either use the better approximation
$\mathrm{TS}_2$ or use the faster TS$_1$ but with
lower thresholds for narrower time windows.

In summary, the approximate TS estimators work well for detection.  To
facilitate comparison of results, we henceforth use the TS$_2$
statistic.  Thus, from the
simulations yielding Figure \ref{fig:tsnull}, we can easily estimate
trigger thresholds with a known false positive rate.  Given the 10$^7$
simulations, we choose a chance false probability of 10$^{-6}$,
corresponding to a value of TS$_2=29.6$ and $\sigma_2=4.7$ and a false
positive rate of about 1 per day assuming a minimum $\delta t$ of
64\,ms. 
In general, these trigger thresholds could be sharpened somewhat with
operational constraints, e.g.  discarding the TS from pixels where the
sky is occulted by the earth.  We also note that these simulations yield a
trigger threshold very close to the on-board GBM trigger
($\sigma_2=4.8$).

Finally, we note here that these simulations demonstrate the
capability of the algorithm to test the null hypothesis, i.e. if the
data consistent with a slowly-varying background.  It will in general
then detect a wide variety of transient signals, including both pulses
of charged particles and \textit{bona fide} GRBs.  Classifying the
sources of transients is a post-detection step, which is not the focus
of this work.  However, in \S\ref{sec:impl} and \S\ref{sec:realworld},
we show that the TS itself can be used for preliminary classification.

\subsection{Transient Sensitivity}

We can now determine the detection threshold by simulating many
realizations of the alternative hypothesis over a range of fluxes to
estimate the fraction yielding a value of TS$_2$ ($\sigma_2$)
above the trigger threshold.  We adopt two windows, $\delta t=$64\,ms
and 1024\,ms, which are of interest for detection of short GRBs and
which facilitate comparison with the catalog of GBM GRBs
\citep{vonKienlin20}.

\begin{figure}
\centering
  \includegraphics[angle=0,width=0.98\linewidth]{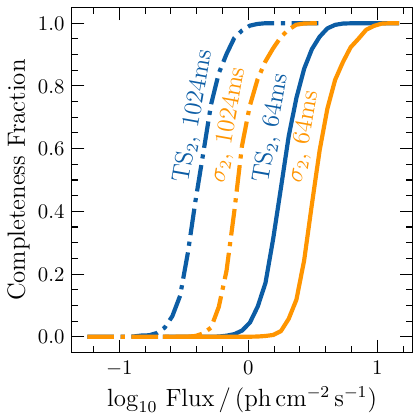}
\caption{\label{fig:roc}The fraction of simulated bursts producing a
  test statistic that surpasses the thresholds discussed in the main
  text, for two time windows, as a function of the 50--300\,keV
  flux of the simulated burst.}
\end{figure}



The results of these simulations are shown in Figure \ref{fig:roc}.
If we take 50\% completeness for a detection threshold, we find values
for 64ms of 1.8\,ph\,cm$^{-2}$\,s$^{-1}$ for TS$_2$ and
3.3\,ph\,cm$^{-2}$\,s$^{-1}$ for $\sigma_2$.  At 1024\,ms, the
thresholds are 0.42 and 0.83\,ph\,cm$^{-2}$\,s$^{-1}$, respectively.
In summary, the ML algorithm is almost twice as sensitive as the
rate-based trigger.


Using the GBM catalog, an instrumental threshold can be estimated from
an observed turnover in the 64-ms GRB fluxes at
$\sim$3\,ph\,cm$^{-2}$\,s$^{-1}$,
in good agreement with our estimate of the $\sigma_2$ threshold.
However, we note that the real instrument samples a much wider range
of background rates and, while its sensitivity peaks in the equatorial
plane (instrumental coordinates), this plane generally intersects the
earth because the rocking profile exposes less sensitive aspects
of the instrument to the sky.  We have not attempted to capture such
details.  Further, these simple simulations assume a top- hat shape
burst that is perfectly aligned with the time window used to extract
data, and so in general realistic thresholds will be higher.  However,
these simplifications are common to both algorithms, so we conclude
that the use of ML methods could substantially improve the onboard
trigger performance of an instrument like GBM.

\subsection{Real-time Localization}

The localization power of an instrument like GBM
or Glowbug depends on the relative brightness of the transient and how
much contrast in count rate between detector elements the instrument
design provides.  The 482 pixels used for the simulations and for the
reference implementation below provide a resolution of about
9$^{\circ}$, which is adequate for estimating the position of
transients near threshold but too coarse for precise localization.
Here, we assess the reliability of coarse positions that result from
the detection algorithm and that might be used in low-latency alerts.
We defer discussion of refined, post-detection localization to
\S\ref{sec:localization}.

If the templates used for detection exactly match the detected burst
spectrum, then the maximum-TS pixel should correspond to the true
(simulated) direction up to the limits of statistical precision.
However, our design mimics real-world applications, with a mismatch
between the true spectrum and the templates, and so the inferred
positions will be biased.

We determined the size of the bias by simulating 10,000 bright
bursts with a 50--300\,keV flux of 10\,ph\,cm$^{-2}$\,s$^{-1}$ and a
duration of 1.024\,s, producing a typical TS$_{\mathrm{e}}$ of 7000.
(NB that for localization, it is critical to use the exact TS.)  For
each simulation, there are three position estimates, i.e. the pixels
that maximize the likelihood for each G20 template.  In general,
these differ, with only 6.4\% of the bursts yielding the same coarse
position for all three templates.  However, by using the
pixel/template combination that maximizes the total TS, we recovered
the correct position 97\% of the time.  Of the remaining 3\% of bursts
with incorrect positions (a substantial systematic error of
$\geq$$9^{\circ}$), more than 90\% were from bursts simulated with
the template that is mid-way between the normal and the hard G20
templatete.  From Figure \ref{fig:models}, it is apparent that the
``distance'' between these templates is larger than between the normal
and soft templates.

Thus, it is clear that the problem of localization cannot be separated
from the question of the true spectrum, and using the wrong template
can generate tremendous ($>$10$^\circ$) errors.  On the other hand,
even with just three templates, the coarse positions from the
detection algorithm are correct almost all the time, suggesting the
prescription can be used ``out of the box'' for GRB direction finding
and for seeding of follow-up localization.

Near-threshold GRBs will in general be poorly localized, but
it is also of interest if the real-time localization of such faint
GRBs is also accurate.  Using the same approach, but adjusting the
source rate to produce a mean TS of about 40, we determined the
distribution of the TS difference between the TS-maximizing pixel and
the pixel of the true position.  We found that $\delta\mathrm{TS}<5.99$
about 92\% of the time, in reasonable agreement with the expectation of 95\%.  Thus, probability maps for faint-GRB localization should be reliable.

\section{Reference Implementation}
\label{sec:impl}

Here, we present a reference implementation (RI) of the ML approach
presented in \S\ref{sec:like} and whose idealized scientific
performance was estimated in \S\ref{sec:sims}.  The RI is designed to
work as a real-time transient detection algorithm on $\gamma$-ray
sensors and to allow real-world computational and scientific
performance evaluations.  Parameters governing some aspects of the RI
are listed in Table \ref{tab:params}; they are fully customizable, but
the specific values used in benchmarking are listed.

The base implementation is in C, and a parallel implementation in
Python allows rapid prototyping and testing of features.  The C
implementation---with additional instrument-specific data handling,
command and control, and housekeeping layers---also provides the burst
detection algorithm for the Glowbug flight software.


\begin{table*}
\centering
\begin{tabular}{l | l | l | l }
\hline
 Name & Symbol & Description & RI Value \\
\hline
\hline
 Detector elements & $\ndet$ & Number of independent detectors & 12 \\
 Energy channels & $\nchan$ & Number of energy channels & 8 \\
 Transient spectra & $\ntmpl$ & Number of spectral templates used for
  search & 3 \\
 Incident directions & $\npix$ & Number of spatial pixels/directions to 
  search & 482 \\
 Input timescale & $\tmin$ & Update timescale for detection
 algorithm & 32\,ms \\
  Number of timescales & $\nf$ & Hierarchical streams & 7 (64\,ms--4096\,ms) \\
  Background timescale & $\tb$ & Update timescale for background model & 1.024\,ms ($32\times\tmin$)\\
  Background samples & $\nb$ & Number of samples used to
  estimate background & 120 / 60 / 30\\
\hline
\end{tabular} 
\caption{\label{tab:params}Parameters, all of which are adjustable,
  used in the reference implementation (RI) of the algorithm.  The
  particular values used in the RI are given in the rightmost column.}
\end{table*}

\subsection{Data Synchronization and Aggregation}

For the RI, we abstract away the frontend electronics and assume as
input a stream of individual events comprising a detector element
index, an energy channel index, and an absolute timestamp.  For our RI
with 12 detectors, these take the form of a 4-byte word that provides
a 1-ms resolution timestamp and an 8-channel energy resolution.  We
further assume that pseudo-events triggered by clock pulses
(PPS, or pulse-per-second) are provided in the datastream from the
onboard clock.

We make no assumptions about raw data ordering, so the first step is
to synchronize the data over detector elements and bin it into uniform
time bins.  Data are discarded until the first PPS is received, which
determines the epochs.  Data are stored in a circular buffer with
resolution $\tmin$ (32\,ms), so a bitshift efficiently converts
epoch-subtracted timestamps into buffer indices.  Data are
synchronized by tracking the detector whose last delivered event is
the oldest, $t_{\mathrm{synch},n}$.  When that detector next delivers
data, a new oldest detector is identified, whose latest event is
$t_{\mathrm{synch},n+1}$.   By construction, all detectors have
delivered at least one event after $t_{\mathrm{synch},n+1}$, so all
bins with $t_{\mathrm{synch},n}<=t<t_{\mathrm{synch},n+1}$ are
synchronized and the buffer indicates these samples are available to
downstream processing.  The output of this procedure is a uniform time
series of counts spectra similar to the \texttt{CTIME} data product of
GBM.

It is necessary to consider transients with a wide range of durations,
and we adopt the same procedure used onboard GBM, aggregating the data
into hierarchical streams that differ in time resolution by
2$\times$.  Each stream has two phases.  Thus, when each new $\tmin$
(32-ms) sample arrives, it is combined with the preceding $\tmin$
sample to update one phase of the $2\tmin$ (64-ms) resolution stream.
It is combined with the next $\tmin$ sample to arrive to update the
second phase.  These two phases feed a $4\tmin$ stream, which
updates every $2\tmin$, etc.  The efficient aggregation means that the
longest timescales searched for transients are set not by
computational constraint but by movement of the spacecraft and by
confusion with the time-variable background.

The RI is tickless: whenever new events arrive, they are written into
the synchronizing buffer, which will produce 0, 1, or more
synchronized, $\tmin$ samples depending on the specifics of the
detector readout scheme.  As soon as any samples are available, they
are fed into the hierarchical summing and the transient search is run
on any updated phases.  (So in the RI, the 64\,ms-timescale search
runs every 32\,ms, etc.) Tasks with an inherent timescale are tied to
particular streams.  E.g., the background mode updates every
$\tau_b=1.024$\,s by tapping one phase of the 1.024\,s stream.

We use a 32-bit floating point format for all data, including counts,
in the RI.  Doing so simplifies data structures, and the memory
requirements for the buffer are quite modest, $<$1\,MiB per minute of
data.

Finally, we briefly consider the computational requirements for the
synchronizing buffer, which is the only component of the RI which
depends directly on the event rate.  In general, operations will
depend on the specific data format.  In the RI, there are about 10
bitwise operations used to extract and test portions of the 4-byte
word, and three integer arithmetic operations (including a modulo) to
determine the buffer index to increment.  For a typical scintillator,
the maximum event rate is set by deadtime and is likely to be
$<$$10^5$\,Hz.  For ten such detector elements operating at the
maximum rate---an extraordinary circumstance---processing the event data will require $10^7$\,s$^{-1}$
integer operations.  This rate is less than the
$\approx$2$\times10^{7}$ floating point operations required for
transient searches (\S\ref{sec:tscomp} and \S\ref{sec:benchmark}).


\begin{table*}
\centering
\begin{tabular}{l | l | l | l }
\hline
Memory\\
\hline
  Name & Complexity & Reference  & MiB \\
\hline
\hline
  Synchronizing buffer & $\nchan \ndet
  \times\frac{\mathrm{t_{buff}}}{\tmin}$ &
  $12\times8\times\frac{600\,\mathrm{s}}{32\,\mathrm{ms}}$ & 6.90 \\
  Templates & $\ntmpl \npix \nchan \ndet$
  & $3\times482\times8\times12$ & 0.53 \\
\hline
Floating Point Ops\\
\hline
  Name & Complexity & Reference  & 10$^6$ OP\,s$^{-1}$ \\
\hline
\hline
  TS$_1$, $+$ & $\mathcal{L}_{\nf}\ntmpl\npix(\nchan\ndet+1)/\tmin)$ &
  $3.97\times 3\times482\times8\times12/32\,\mathrm{ms}$ & 17.4 \\
  TS$_1$, $\times$ & $\mathcal{L}_{\nf}\ntmpl\npix(\nchan\ndet+1)/\tmin$ &
  $3.97\times 3\times482\times8\times12/32\,\mathrm{ms}$ & 17.4 \\
  TS$_1$, $\div$ & $\mathcal{L}_{\nf}\ntmpl\npix/\tmin$ &
  $3.97\times 3\times482/32\,\mathrm{ms}$ & 0.18 \\
  Bkg. update, $\times$& $\ntmpl\npix\nchan\ndet/\tb$ &
  $3\times482\times8\times12/1024\,\mathrm{ms}$ & 0.14 \\
\hline
\end{tabular} 
\caption{\label{tab:comp}Computational requirements in terms of memory
and floating point operations.  The second column indicates the
  theoretical complexity and, where applicable, the memory layout
  of arrays.  The third column gives the specific values for the
  reference implementation, and the fourth either the required storage in bytes
  (all storage and operations are implemented with 32-bit precision
  floating point) or the required number of floating point
  operations.}
\end{table*}

\subsection{Background Estimation}

In \S\ref{sec:sims} we assumed perfect knowledge of the background,
but in application it must be estimated.  The primary challenge is the
orbital variation in both the incident particle background and
internal background from activation, especially from passages through
the South Atlantic Anomaly (SAA), a region with a high flux
of energetic particles.  The variations further depend on the orbital
precession phase, the solar cycle, and solar flaring activity.

Ideally, background variations could be predicted with parametric
models \citep{Biltzinger20} or from archival data from earlier orbits.
However, with high background count rates, even small errors in the
background model produce spurious or missed transient detections.
Thus, for the RI we estimate the background with a simple, robust
moving average.

Specifically, we collect $\mathrm{N_b}$ (120) samples of data
from the $\tau_{\mathrm{b}}$ stream (1.024\,s) and from this
estimate the mean and slope of the count rate for each
detector and channel.  To avoid overlap of the background samples with
those being searched for transients, we project the background
estimate forward in time, typically about 4\,s.

This linear model can only capture some of the true background
variations, so we also attempt to determine if the estimate is
consistent with the data.  Of particular concern are (1) intervals
when the instrument is entering or exiting the SAA, where the
background rate may change nonlinearly over the time $\nb \tb$, and
(2) rapid ``spikes'' in background due to incident particles, e.g. 
particle precipitation events.  To mitigate the first
issue, we invalidate the background estimate if the magnitude of
the observed slope exceeds a maximum per-channel value.  To
mitigate spikes, we observe the residuals of the samples in the
background window.  If the model is adequate, then these residuals
should follow an approximately normal distribution with variance
equal to the typical count rate.  We test for normality
using a kurtosis-only version of the D'Agostino test
\citep{DAgostino90}, and if there is substantial unmodeled rate
variation (exceeding the D'Agostino test threshold), the background
model is invalidated.

It is often too restrictive to require linear background variation
over $\nb\tb=123$\,s, so the RI additionally implements nested models
operating on $\nb/2$ and $\nb/4$ samples.  If the longest window is
invalid, the shorter windows are considered, with larger thresholds
for the time derivative and D'Agostino test.  Whenever the
background model is in steady state, the preferred, longest window is
used.  If a (short) rate excursion occurs, all three windows will fail the
check, but within $\nb/4$ samples, the spike will leave the shortest
window, which will furnish a new background estimate.
$\nb/4$ samples later, the next longest window ($\nb/2$) will be
adopted, and finally the spike will exit the full window and steady
state is restored.  This approach preserves 75\% of the exposure
compared to a single background window.


In general, most parameters in the RI are adjustable but have been
tuned for the orbit and altitude of GBM.  We show examples of the
background estimator in operation in \S\ref{sec:realworld}.  We expect
that this approach will also be suitable for orbits of instruments
such as Glowbug (inclination 52$^\circ$) with suitable parameter
updates, but may require additional features to capture and mitigate
background variations in the auroral zones.

Because the computation associated with the background
model update has complexity $\mathcal{O}(\ndet\nchan)$, it is
negligible compared to the evaluation of TS.

\subsection{TS Computation}
\label{sec:tscomp}

Whenever a new sample becomes available in one of the time-averaged streams,
we search for a transient by evaluating TS in a four-dimensional
loop over $\ntmpl$ templates (index $k$), $\npix$ spatial directions
($l$), $\nchan$ energy channels ($i$), and $\ndet$ detector elements
($j$).  From Equations \ref{eq:ts1} and \ref{eq:ts2}, the innermost
loops are an evaluation of an inner product between the observed
counts and the predicted signal-to-background ratio:
\begin{align}
  NT_{kl} =& \sum_{i=1}^{\nchan}\sum_{j=1}^{\ndet}
  c_{ij} F_{klij}/b_{ij} \equiv c_{ij} t_{klij}\\
  NT^2_{kl} =& \sum_{i=1}^{\nchan}\sum_{j=1}^{\ndet}
  (c_{ij} t_{klij}) t_{klij}\\
  NT^3_{kl} =& \sum_{i=1}^{\nchan}\sum_{j=1}^{\ndet}
  (c_{ij} t^2_{klij}) t_{klij}.
\end{align}
In the innermost loop, each moment requires a single multiply and add,
and so from Equation \ref{eq:ts1} it is apparent that evaluating
TS$_1$ thus requires $2\ntmpl\npix(\nchan\ndet+1)$ adds and multiplies
and $\ntmpl\npix$ divisions.  From Equation \ref{eq:ts2}, evaluating
TS$_2$ requires an additional $\ntmpl\npix(\nchan\ndet+1)$ adds and
$\ntmpl\npix(\nchan\ndet+3)$ multiplies.  Some architectures are
likely to support the sequential multiplies, adds, and assignments via
fused operators.  Up to negligible factors, TS$_2$ requires about 50\%
more multiplies and adds than TS$_1$.  In the RI, the maximum TS for
each template is recorded, requiring $\ntmpl\npix$ floating point
comparisons, likely also to be negligible except on hardware where
comparisons and jumps are anomalously slow.

It is convenient to scale these computational requirements per input
sample.  In the RI there are 7 streams (64\,ms, 128\,ms,
\ldots, 4096\,ms), so the amortized computations per input are
$\mathcal{L}_{N_F}\equiv\sum_{i=1}^\nf 2^{1-i}\approx1.98\times$ those
required to evaluate 64\,ms stream.  Thus, for
evaluation of TS$_1$, for each $\tmin$ (32\,ms) input sample, there are
(amortized) $3.97\ntmpl\npix(\nchan\ndet+1)$ adds and multiplies,
$3.97\ntmpl\npix$ divides, and $\ntmpl\npix\nchan\ndet\tmin/\tb$
multiplies for the background update.

The most substantial ancillary computation is the pre-computation of
$t_{klij}$, requiring $\ntmpl\npix\nchan\ndet$ multiplies every
$\tau_b$\,s.  When the background model is ``bad'', $t$ is set to 0,
producing 0 for all evaluations of TS.

In addition to the standard, template-based TS computation, we have
determined that a per-channel TS can help to identify transients due
to spikes in the particle background.  The most useful of these are
the lowest channel, appropriate for soft electrons, and the highest
channel, typically for rapidly rising proton rates on entering the
SAA.  The per-channel TS computation is very similar to that outlined
above, but requires additional overhead in the ``channel'' loop that
is surprisingly expensive.  In the RI, we compute the per-channel TS
for all $\nchan$ channels, but suggest it should be tuned to
application.  We have further chosen a memory layout that simplifies
programming logic (there is always a sum over $\ndet$ in the innermost
loop), but swapping the innermost two dimensions may be favorable on
some architectures.

\subsection{Triggering}

In general, any TS value---from any stream---that
surpasses a pre-defined threshold value could be considered a trigger.
However, it may be advantageous to prefer longer time windows, which
generally improve statistical precision, have reduced trials factors,
and could provide better seed localizations for follow-up analysis.
Further, the use of per-channel TS can help to reject background
variations, and we have found that this is more effective with the
improved precision of longer time windows.

Thus we distinguish a local trigger---any time
the TS surpasses the threshold value---from a global trigger, which
is evaluated only on the longest averaging timescale and which can 
change the instrument mode to a triggered state.

\begin{table}
\centering
\begin{tabular}{l | l | l | r }
\hline
Name & Processor & Clock Speed & GFLOPs \\
\hline \hline
Xeon & Intel Xeon W-2123 & 3.6\,GHz & 13.0 \\
Rpi4 & ARM Cortex A72 & 1.5\,GHz & 1.7 \\
Q8   & ARM Cortex A53 & 1.2\,GHz & 0.5 \\
  \hline
\end{tabular} 
  \caption{\label{tab:bench_systs}Brief specifications for benchmark
  systems.}
\end{table}

\begin{table*}
\centering
  \hspace{-2.5cm}
\begin{tabular}{l | r | r | r | r | r | r }
\hline
& Xeon & Xeon & RPi4 & RPi4 & Q8 & Q8 \\
\hline
Task&Time (s)&Fractional & Time (s) & Fractional & Time (s) &
  Fractional \\
\hline
\hline
Event processing & 0.26 & $4.8\times10^{-5}$ & 0.54 &
  $1.0\times10^{-4}$ & -- & -- \\
Bkg model and update & 0.29 & $5.4\times10^{-5}$ & 0.99 & $1.8\times10^{-4}$  & -- & -- \\
TS$_1$ & 9.36 & $1.7\times10^{-3}$ & 61.11 & $1.1\times10^{-2}$  & 148
  & $2.7\times10^{-2}$ \\
TS$_2$ & 11.92 & $2.2\times10^{-3}$ & 85.35 & $1.6\times10^{-2}$  &
  183 & $3.5\times10^{-2}$ \\
Per-channel TS$_1$ & 15.34 & $2.8\times10^{-3}$ & 38.54 &
  $7.1\times10^{-3}$ & 98 & $1.8\times10^{-2}$  \\
Per-channel TS$_2$ & 17.41 & $3.2\times10^{-3}$ & 56.56 &
  $1.0\times10^{-2}$  & 115 & $2.1\times10^{-2}$ \\
Other & $<$0.30 & $5.6\times10^{-5}$ & $<$1.48 & $2.7\times10^{-4}$ & -- & -- \\
\hline
  Total & 29.99 & $5.6\times10^{-3}$ & 145.52 & $2.7\times10^{-2}$ &
  298 & $5.5\times10^{-2}$ \\
\hline
\end{tabular} 
  \caption{\label{tab:bench}Execution times required to process
  5400\,s of data.  The ``Time'' columns indicate the absolute
  processing time, while the ``Fractional'' columns give the ratio of
  the processing time to the total time (5400\,s).  In all cases, the
  TS computation dominates the run time, and in all cases, the total
  fractional processing time is $\ll$1, indicating suitability for
  real-time operation.  Note that TS$_2$ times are inclusive of the time required to compute TS$_1$.}
\end{table*}

We assess the presence of a global trigger as follows.  For each
update of the longest timescale (4096\,ms, every 2048\,ms), we check
for local triggers in all of the shorter windows that overlap the
global window and select the one with the highest TS.  E.g. in the
event of a very short GRB (say 160\,ms), then the 128\,ms window
should be selected, while a long GRB (say 20\,s) should produce the
highest TS in the 4096\,ms window.

We select the optimal window for both the template TS and for the
per-channel TS from the lowest- and highest-energy channels.  If these
per-channel TS values are comparable or greater than the template TS,
it indicates the transient consists primarily of very soft or very
hard particles and is thus likely associated with charged particle
background.  In the RI, a global trigger is initiated when the
template TS exceeds both the soft and hard-channel TS by a factor of 1.3.

\subsection{GBM Data Playback}

We use GBM data both for the benchmarks reported below and in
\S\ref{sec:realworld} to test the scientific performance of the transient
search algorithm.  To do this, we break the archival data set up into
1-day intervals, and further divide this data set into ``orbits'',
which are determined either when GBM enters the SAA or when 90
minutes have elapsed.

For each day, we begin with the archival \texttt{TTE} format data and
take the intersection of the Good Time Intervals for each detector.
We re-channelize the data from the original 256 to approximately match the 8 desired channel boundaries.  We insert PPS
events into the data stream at each 1-s boundary, then group data into
packets of up to 250 events, converting each event timestamp, channel, and
PPS flag into the 4-word format used by the RI (and by Glowbug).
Finally, we order the packets according to the timestamp of the last
event in them, thus emulating the staggered data delivery expected in
real-time application.

\subsection{Benchmarking}
\label{sec:benchmark}

To compare with the floating point complexity estimates made above,
and to provide realistic performance expectations, we benchmark the RI
on several distinct platforms.  Specifically, we play back the first
5400\,s of GBM data acquired on 9 Feb 2014, which has an average
summed event rate (within the energy range) of 7.3\,kHz.
For each platform, we consider the
following scenarios:
\begin{itemize}
\item The RI is run with the TS computation disabled.  This allows
  estimation of overhead and the event processing rate, further
    isolated with the use of \texttt{gprof}.
\item The RI with template-based TS$_1$ computation enabled.
\item The RI with template-based TS$_2$ computation enabled.
\item The RI with template and channel-based TS$_1$ computation enabled.
\item The RI with template and channel-based TS$_2$ computation enabled.
\end{itemize}
Together, these benchmarks provide a full indication of the
computational requirements for real-time use on an arbitrary platform.

For each hardware platform, we adopt the compiler command \texttt{gcc
-pg -g -O3 -ftree-vectorize -ffast-math -march=native -std=c11}.
Further, we use the ``linpack'' benchmark to estimate a characteristic
single-precision, single-threaded floating point performance (Table
\ref{tab:bench_systs}).




There are exactly 168718 32\,ms samples processed in the benchmark
run, and so from Table \ref{tab:comp}, we expect the TS$_1$
calculation to require, in the inner loop, $9.3\times10^{10}$ each of
adds and floats.  For the Xeon, with estimated single-precision
floating point rate of 13\,GFLOPs, the estimated required time is
14.3\,s, about 50\% higher than the observed value, indicating some
effectiveness of SIMD vectorization of loops.  We expect that an
implementation with hard-coded loop invariants ($\ndet$, $\nchan$,
etc. could all be fixed in a flight system) would produce more
aggressively optimized code, as would manual implementation of the
inner loop assembly code.  However, the code is already very fast on a
Xeon, requiring $<$1\% processor utilization on a single core to
process the data in real-time equivalent.

The Cortex A72 on the Raspberry Pi 4 is a much less performant
processor, running at half the clock speed and with much smaller pools
of cache and functional units.  As it is also the flight processor for
Glowbug, the performance on this platform is critical.  It completes
the benchmark run about 5$\times$ slower than the Xeon, meaning it is
capable of real-time data processing with $<$3\% processor utilization.

\begin{figure*}
\centering
  \includegraphics[angle=0,width=0.98\linewidth]{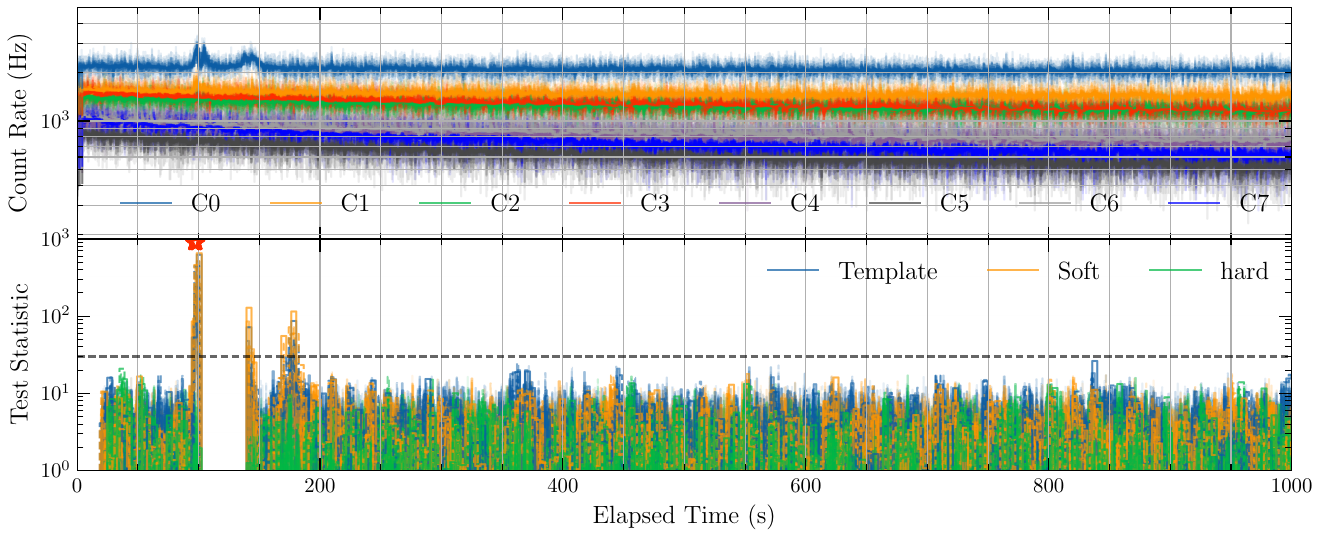}
\caption{\label{fig:day_140208}Example of the application of the
  maximum likelihood algorithm to archival GBM data from 08 Feb
  2014 beginning at MET 413575773.952.  The top panel shows the count rate summed over all 12 NaI
  detectors and grouped into the 8 energy channels  listed in Table
  \ref{tab:channels}.  The rate decreases as energy increases.
  All streams are shown overlaid, with
  the faster ones using a higher degree of transparency.
  The top panel gives the maximum template TS (over 482 positions and
  the three G20 templates)
  Also shown are the soft-channel and hard-channel TS values.
  There is a clear transient, generated by
  incident electrons, which appears primarily in the soft channel C0,
  though the initial pulse is also visible in channel C1 ($>$50\,keV).
  These particles generate a local trigger (unfilled red star), but because the
  soft-channel TS is comparable to the template TS, no global trigger
  is issued.  The transient causes the background model to be flagged
  as invalid while the initial pulse lies in the background window;
  during these intervals, TS$=$0.  The remainder of the 1\,ks of data
  show a steady count rate and correspondingly a TS well below
  threshold (indicated here by the black dashed line at TS$=$30).}
\end{figure*}

Finally, we consider a processor designed specifically for space
applications, the radiation-hardened Xiphos Q8.  Its system-on-a-chip
includes a 4-core Cortex A53 as well as an FPGA fabric, though we
consider only evaluation of the CPU.  We built a custom \texttt{yocto}
image with czmq support and cross-compiled the RI.  Evaluation of the
full suite at real-time requires $<$6\% utilization of a single core, and
the scaling is in good agreement with the relative difference in
floating point performance compared to the A72.

In summary, the RI implementation supports real-time ML-based burst
detection on processors likely to be selected for a deployment in
space, both at low- and high-cost levels.

\section{Real-world Detection Performance}
\label{sec:realworld}

\begin{figure*}
\centering
  \includegraphics[angle=0,width=0.98\linewidth]{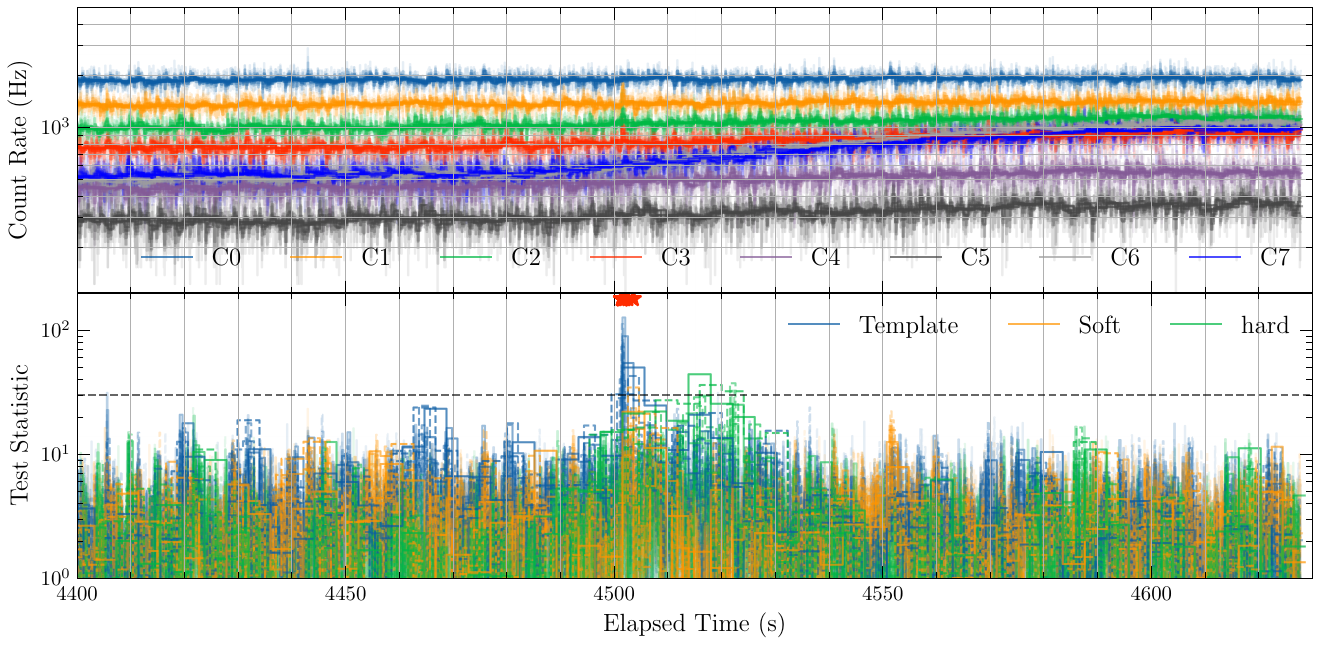}
  \caption{\label{fig:day_170817}As Figure \ref{fig:day_140208}, but
  showing data beignning at MET 524661967.952, a few minutes
  surrounding GRB~170817A.  The GRB is clearly detected and generates
  a global trigger (filled red star).  The maximum observed TS is 107,
  occurring in the 256\,ms streams.  Since the GRB is relatively faint,
  the background model remains valid.}
\end{figure*}

Here we show two examples of applying the RI implementation to
archival GBM data.  The first, in Figure \ref{fig:day_140208}, shows
1000\,ks of data containing a pulse of incident electrons on top of an
otherwise slow evolution of the background rates.  This example
illustrates the computation of TS.  Several local triggers are
generated in the faster streams because the TS is a random variable
and suffers more noise in short time windows.  On the other hand, in the
longer windows, the template TS is always lower than the soft-channel
TS.  These local triggers are vetoed and indicated by the unfilled red
stars.  Extensive testing has shown this prescription is an effective
means of identifying such particle transients, even excluding soft
events to mimic an instrument with a higher threshold (30--50\,keV) compared to GBM.
(The $\sim$10\,keV threshold for GBM makes electron events
particularly obvious.) Following the arrival of the first bright
electron pulse, the background model begins to fail the
self-consistency check and is flagged as invalid, which automatically
sets TS to 0.  Following the initial pulse, the background model is
re-established.  The fainter, slower electron pulses only generate
large values of TS in the slow streams and are always vetoed.

The second example shows the ML algorithm applied to GRB~170817A.  It
generates a global trigger (filled star) with a maximum TS$_2$ of 118 in
the 256\,ms window.  \citet{Goldstein17} note that the GRB generated
significant detections in three detectors, and that the strongest
signal was achieved with the 512\,ms filter, with the second-highest
significance yielding a $\sigma_2$ of 6.3$\sigma$.  Consequently GBM would have detected
GRB~170817A onboard if it were up to 40\% fainter.  On the other hand, with a
reasonable threshold of TS$=30$, the ML approach would have allowed GBM
to detect GRB~170817A if it were 4.3$\times$ fainter!  Alternatively,
it could have detected it at twice the distance, increasing the
rate of such GRBs by about 8.  This claim is somewhat conservative
because the RI only uses data $>$30\,keV.

\begin{figure*}
\centering
  \includegraphics[angle=0,width=0.98\linewidth]{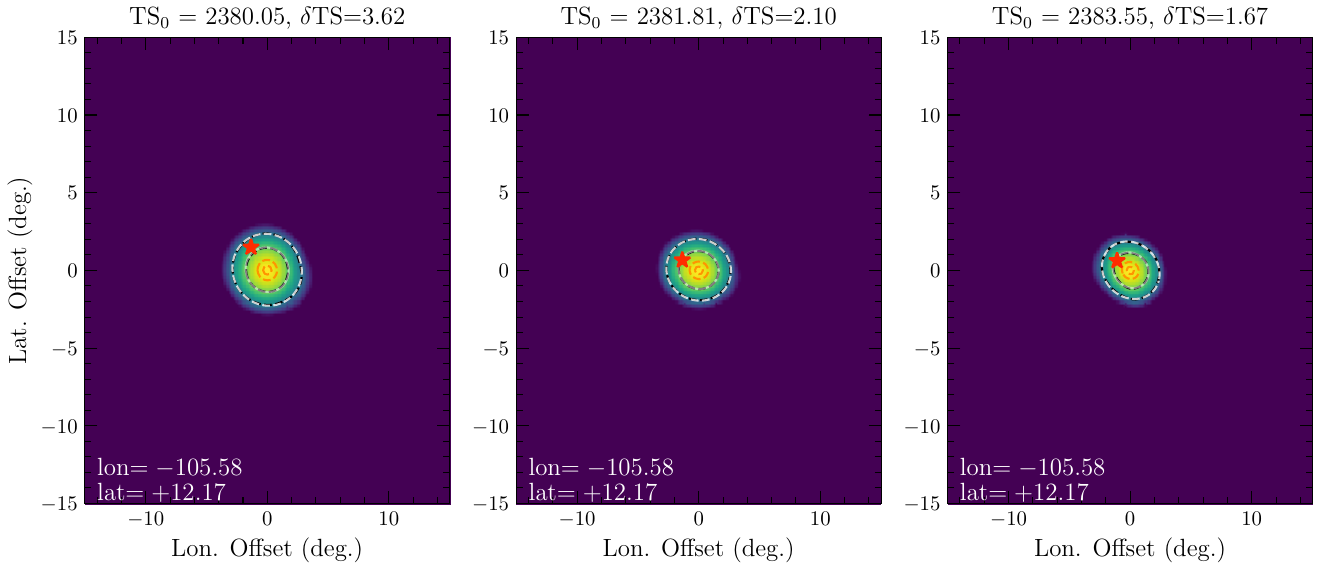}
\caption{\label{fig:tsmap_normal}The TS map for three resolutions of
  the RM, using 192 (left), 642 (center) and 2562 (right)
  pixels.  The TS is evaluated on a 15$^{\circ}$$\times$15$^{\circ}$
  grid in Cartesian projection around the position that maximizes the
  overall TS.  The true (simulated) position of the transient appears
  as a red star.  The black contours indicate TS drops of 5.99 and
  2.30, corresponding to 95\% and 68\% confidence regions.  The red
  contours indicate TS drops of 0.5 and 0.1, simply for reference.
  The white contours indicate the 68\% and 95\% confidence regions for
  the error ellipse that best fits the TS surface.  The simulated
  transient position is annotated at lower left, and the resulting
  maximum TS and the TS difference between the simulated position and
  the best-fit position ($\delta$Ts) are annotated above each panel.
  }
\end{figure*}

\begin{figure*}
\centering
  \includegraphics[angle=0,width=0.98\linewidth]{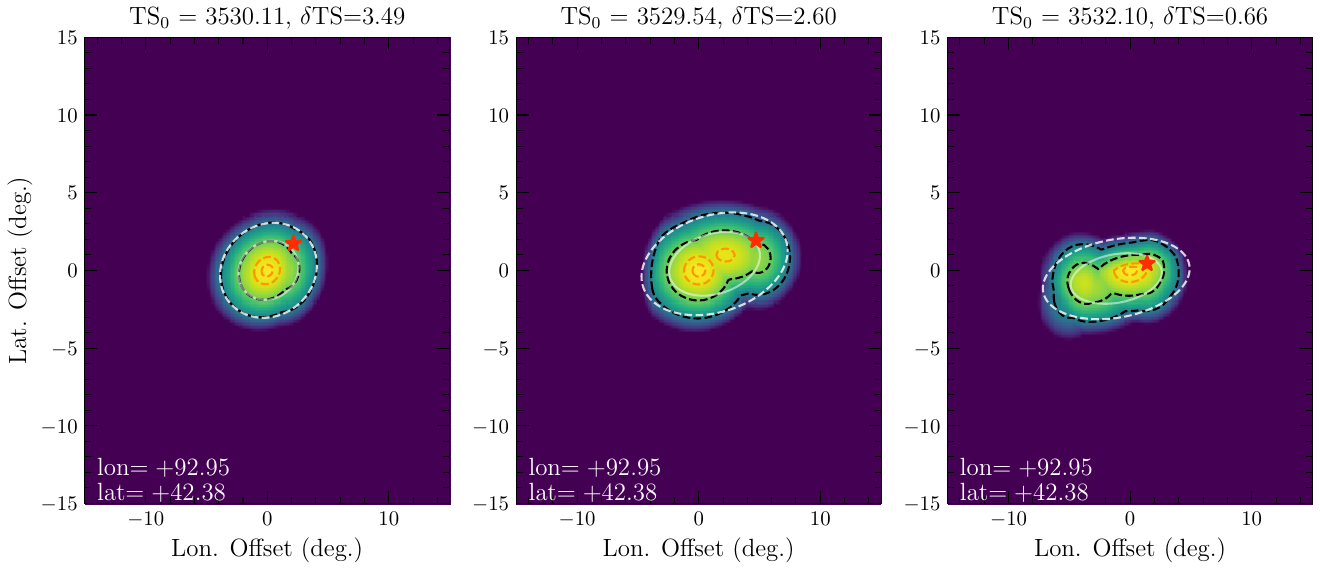}
  \caption{\label{fig:tsmap_anti}As Figure \ref{fig:tsmap_normal}, but
  showing a case where the higher spatial resolution RM reveals a more
  complicated TS map.}
\end{figure*}

\begin{figure*}
\centering
  \includegraphics[angle=0,width=0.98\linewidth]{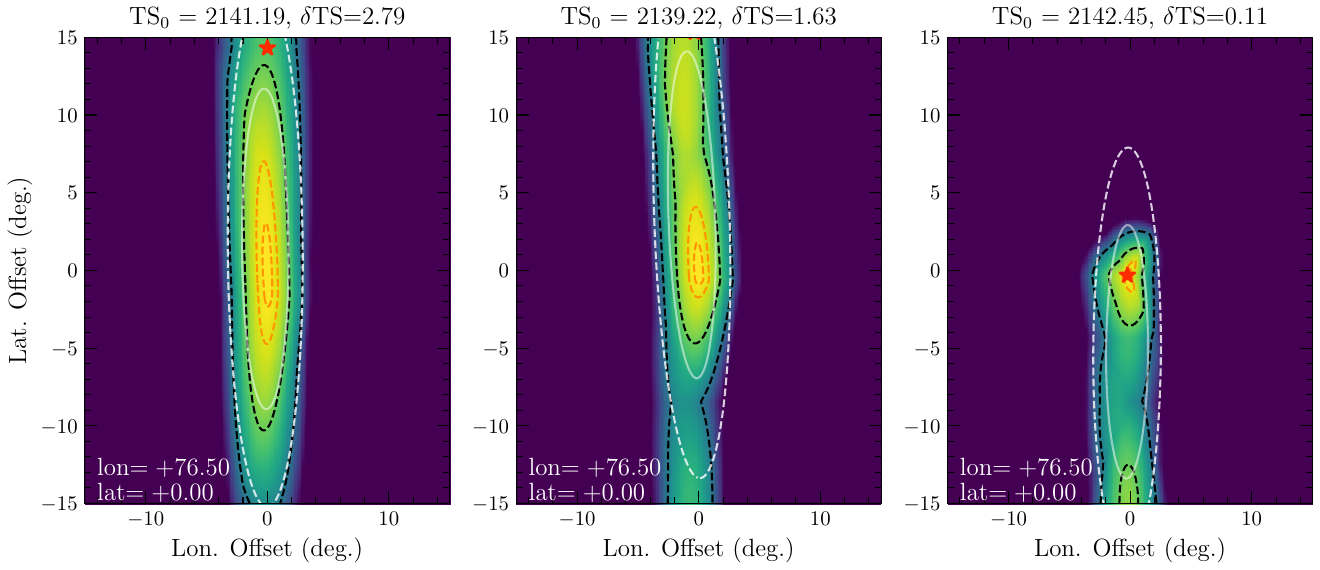}
  \caption{\label{fig:tsmap_lowlat}As Figure \ref{fig:tsmap_normal},
  but for a transient incident from the southern hemisphere, producing
  elongated and complex uncertainty regions.}
\end{figure*}

\section{Precise Localization}
\label{sec:localization}

As discussed and validated above, ML-based detection automatically
provides coarse localizations.  In principle, finer real-time localizations
could be obtained simply by searching over more directions, but
degree-scale precision is likely too expensive for low-power
processors.  Instead, these coarse localizations are appropriate as
seeds for a follow-up analysis that produces a refined position
estimate.

The RI does not have such a follow-up capability, but the additional
computational complexity is small.  A straightforward approach is to
generate a degree-scale ``TS map'' by evaluating the TS over a grid
that covers the full coarse pixel from the seed location.  Such a map
is sufficient to generate a centroid and confidence region and
requires only $\sim$100 additional TS evaluations, a negligible
increase on those performed by the detection algorithm.

Here, we test how well this procedure might work---and the reliability
of ML localizations generally.  It is well-demonstrated that GBM
localizations---which use ML---suffer from a systematic error of a few
degrees \citep{Connaughton15}.  Such localizations rely on two key
ingredients: the ``true'' spectrum of the burst, and the ``true''
instrument RM to that burst.  The GBM RM is based on Monte Carlo
particle transport simulations \citep{Kippen07,Hoover08} of the
\textit{Fermi} Gamma-ray Space Telescope sampled over 272 incident
directions.  If the mass model used in the simulations is incomplete
or inaccurate, then the RM will also suffer inaccuracies.  On the
other hand, \citet{Berlato19} note the importance of an accurate
spectral template and claim to reduce systematic errors by using using
more degrees of freedom in the spectral model.

We consider a related possibility: the effect of the number of pixels
used in the representation of the RM.  To do this, we used
\texttt{SWORD} \citep{Duvall19} to create a detailed mass model for
Glowbug (instrument only), performed Monte Carlo simulations of an
incident broadband $\gamma$-ray spectrum on Glowbug with the
underlying \texttt{GEANT4} \citep{Allison16} radiation transport
package, and evaluated the RM at three spatial resolutions, using 192,
642, and 2562 incident directions.  We do not consider
atmospheric scattering, which is an important consideration in GRB
localization but which essentially only changes the response matrix.

We then simulated 2000 realizations of data as in
\S\ref{sec:sims} with the following differences: (1) we used the
2562-pixel Glowbug RM instead of the GBM RM; (2) we used only the ``normal'' G20 template (but continue to
select a random incident pixel) with a
50--300\,keV flux of 3\,ph\,cm$^{-2}$\,s$^{-1}$; (3) we simulated
1.024\,s time windows; (4) we restricted the incidence polar angle
to $\cos\theta>0$, i.e. the portions of the sky with the
best Glowbug sensitivity.  These parameters yield a fairly bright burst
relative to the background that can typically be localized with a
precision of a few degrees.  For each simulation, we processed the
data using the three spatial resolutions, specifically (1) determining
the coarse pixel that maximized the TS according to the GRB template;
(2) adaptively refining the localization by spherically, linearly
interpolating the instrument response to predict the counts on a finer
grid; and (3) producing a final 15$^{\circ}$$\times$15$^{\circ}$
TS map on a uniform (Cartesian projection) grid around the
maximum TS position.  NB that unlike as in \S\ref{sec:sims}, we use
the same GRB spectral template for both simulation and evaluating the
TS in order to isolate the systematic effect of RM resolution.

A good localization must deliver accurate estimates of
both the position and its uncertainty, since the latter often governs
whether or not it is possible to follow-up transients with
narrow-field instruments.  The TS is a random field distributed as
$\chi^2_2$, so confidence intervals around the best-fit position can
be estimated as, e.g., the contour by which the TS drops from the
maximum value by 2.30 (68\%) or 5.99 (95\%).  Alternatively, for cases
where the likelihood surface is not particularly gaussian, the
likelihood can be combined with a uniform prior and treated
numerically to derive Bayesian credible intervals.  Thus, we can
assess the quality of the localization by analysis of the TS maps
produced for the three resolution levels.

The TS maps from the simulations can be grouped into several classes.
An example from most common class appears in Figure
\ref{fig:tsmap_normal}.  The transient is well-localized at all three
resolutions, and the uncertainty contours are effectively gaussian, as indicated by
the agreement between the TS map contours with a quadratic fit to the
surface.  Such a result could be accurately encapsulated as a single
position and an uncertainty ellipse.  As the spatial
resolution of the RM increases, the size of the error
ellipse decreases.  (The TS difference between the simulated position
and the best-fit position also decreases.)  As we will see, this is a
general trend.

The second example, Figure \ref{fig:tsmap_anti}, shows a class where
the smoothing in the low-pixel RM generates misleadingly gaussian TS
maps while the high-pixel RM yields a TS map
with a more complicated structure.  In these cases, it is more
suitable to describe the position with the full map.  This more
complex surface includes the true position in the
high-probability region, whereas it is in low-probability region (outside the
95\% contours) in the low-resolution versions.

Finally, we consider a transient incident in the equatorial plane of
Glowbug.  Because the detectors form a half cube, such low-latitude
bursts can be moved ``up'' and ``down'' without changing the predicted
counts very much, and the uncertainty regions thus become narrow
ellipses, as shown in Figure \ref{fig:tsmap_lowlat}.  Here, because
the true count rate varies little over the long ellipse, the
systematic differences between the RMs are magnified, and we see the
high-resolution RM yields a TS map with substantially more structure
and that again better includes the true position.

\begin{figure}
\centering
  \includegraphics[angle=0,width=0.98\linewidth]{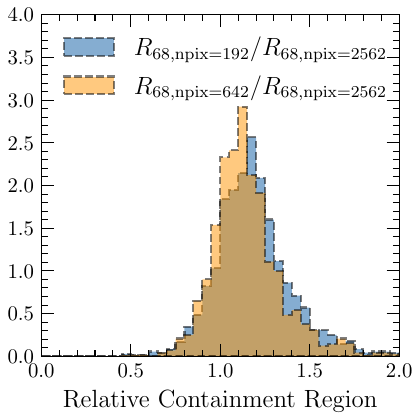}
\caption{\label{fig:loc_comp_r68}The distribution of relative sizes
  of the size of the 68\% confidence region, $R_{68}$, for the two
  low-resolution RMs as compared to the highest-resolution version.  }
\end{figure}

\begin{figure}
\centering
  \includegraphics[angle=0,width=0.98\linewidth]{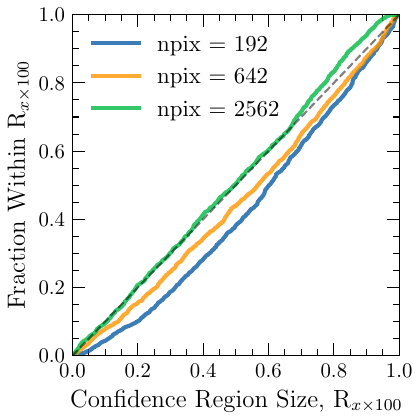}
\caption{\label{fig:loc_comp_pp}The cumulative distribution of the
  offset in position between the simulated and best-fit position,
  scaled by the size of the uncertainty region.  Specifically, this is
  estimated from the cumulative $\chi^2_2$ distribution evaluated for
  the difference in TS between the positions.  If the positions and
  uncertainty regions are accurate, the distribution should be
  linear.}
\end{figure}


To synthesize these results, we consider two key questions: what is
the relative size of the uncertainty regions, and how often is the
true position of the transient within them?  In Figure \ref{fig:loc_comp_r68}, we show the relative
ratios of sizes of $R_{68}$, the region expected to contain the
true transient position 68\% of the time, estimated directly from the
TS map as the summed area of all pixels within
$\delta\mathrm{TS}<2.30$ of the peak.  From this, it is clear that the
confidence regions are larger when using the low-resolution RMs: the
median region is 17\% larger for 192 pixels and 13\% for 642 pixels.
But the tails are heavy, with e.g. 16\% of the regions are 40\% larger
(192 pixels) and 33\% larger; and the worst 5\% of the regions are 61\%
larger and 54\% larger, respectively.

Next, we consider the reliability of the uncertainty regions, which is
best assessed by considering the difference in TS between the
best-fit and simulated positions, $\delta$TS.  The probability to
observe a transient with a given offset is then simply
$p=\int_0^{\delta\mathrm{TS}}\chi^2_2(x)\,dx$.  Then, if the TS
distribution truly follows the $\chi^2_2$ distribution (the
uncertainty estimates are accurate), $p$ should follow a uniform
distribution.  Figure \ref{fig:loc_comp_pp} shows the cumulative
distribution of $p$, from which it is clear that the low-pixel RMs yield
likelihood surfaces that underestimate the uncertainty, particularly
for the 192-pixel version.

All of these effects would be magnified if considering even more
off-axis bursts with $\cos\theta<0$.  In conclusion, we find that it
is critical to use a high-resolution response matrix, with about an
order of magnitude more pixels than are currently in use by, e.g., GBM
analysis tools.  Doing so yields both improved constraints on position
and reduces systematic error stemming from inaccuracies in the TS maps
used to infer positions.  The magnitude of these errors are often
several degrees, in the case of these bright bursts, and thus it is
plausible that this imprecision in the RM, \textbf{along with
potential inaccuracies in the \textit{Fermi} mass model},  contributes
substantially to the observed systematic errors in GBM positions.

\section{Discussion}
\label{sec:discussion}

We have developed maximum likelihood algorithms that are fast enough
to run in real-time on low-power processors.  Their trigger thresholds
can be calibrated with a predictable false-positive rate, and for a
given trigger threshold they deliver a detection threshold about half
that of existing on-board triggers.  In a test with archival GBM
data (GRB~170817A), the sensitivity was improved by more than
$4\times$.

From an instrumental design standpoint, this is a substantial
sensitivity boost.  With thin scintillators, both the signal and
background rates scale approximately linearly with effective area,
$A$, so the signal-to-noise ratio for a given transient only scales as 
$\sqrt{A}$.  Because we have shown that
the computational burden is modest and requires no unusual hardware,
adopting ML algorithms allows an instrument to reach the same
sensitivity as one about four times larger for ``free''.

From a scientific standpoint, it is also a substantial gain.  The
transients of most interest for multimessenger astronomy are
relatively nearby, and thus uniformly distributed through a detection
volume such that $\log N$--$\log S\propto S^{-3/2}$.  This degree of
improvement of sensitivity yields a transient rate that is increased
to $1.8^{\frac{3}{2}}$ to $2.0^{\frac{3}{2}}$ (240--280\%) of the
baseline rate.

We note that with GBM data, this sensitivity improvement can be and
has already been exploited in the ``sub-threshold'' search pipeline
\citep{Blackburn15,Kocevski18}, which is possible post facto because
GBM can downlink every photon event with TDRSS.  (In principle the
CTIME data would be sufficient for ML detection, but the low spectral
resolution would complicate analysis.) Access to such high-bandwidth
telemetry is not guaranteed for future experiments, and in particular
Starburst will not be able to send its full event data to the ground.
However, we have shown it is possible to realize ML techniques with
on-board processing, ensuring that faint transients can be identified
while only downlinking data of interest.

Because the RI presented here only requires $<$10\% of a single core
of low-power processors, we can consider even more ambitious
applications.  Possibilities include searches for ms-scale transients,
such as terrestrial $\gamma$-ray flashes \citep[TGFs,
e.g.][]{Roberts18} or the use of a truly fine spatial grid to provide
real-time near-degree-scale localizations.  At a minimum, the use of
an expanded library of GRB templates relative to the RI could reduce
systematic errors in rapid localization.  Such rapid, accurate
localization would facilitate robotic follow-up observations of
rapidly-fading afterglows.

Our investigation into the robustness of precise burst localization
revealed a surprising dependence on the spatial resolution of the
instrument RM, and we suspect this effect contributes
to the systematic uncertainty of GRB localizations with GBM.
\citet{Goldstein20} find evidence for a two-component model in which
about half of GRBs have a small uncertainty concentrated around
$\sim$2$^{\circ}$, and the other half occupy a long tail with a
typical value of $\sim$4$^{\circ}$.  (Previous analyses of BATSE data \citep{Briggs99} and GBM data \citep{Connaughton15} found similar distributions.)  Both values are consistent with
classes of systematic errors we observed in our analysis with
low-resolution RMs, and we speculate that the two components may
simply result from the proximity of the true GRB position to one of
the RM sampling points.  This idea could be tested with a
higher-resolution RM for GBM.  We used the Glowbug RM for our study
because we had developed substantial machinery for creating mass
models, performing Monte Carlo, and parsing the results into detector
RMs.  The GBM mass model is substantially more complicated (including
the entire \textit{Fermi} spacecraft), and creating a higher
resolution RM is a substantial undertaking.  However, archival and
future GBM data are critical for multi-messenger astronomy, and
obtaining more reliable localizations make it a worthwhile endeavor.

Finally, the relative ease of performing ML detection on-board also
informs GRB experimental design.  In particular, it increases the
value of instruments with good ``geometry factors'', i.e. those that
expose discrete detector elements to different portions of the sky,
like GBM.  Given the rapidly improving availability of affordable
SmallSat spacecraft buses and the development of large-area (heavy)
scintillator detectors, this medium-scale form factor may be the
``sweet spot'' for future networks of $\gamma$-ray sensors.

\begin{acknowledgements}
This work is supported by the Office of Naval Research.
\end{acknowledgements}

\facilities{Fermi}

\bibliographystyle{aasjournal}
\bibliography{sr}


\end{document}